\documentclass[a4paper]{IEEEtran}
\IEEEoverridecommandlockouts
\usepackage[utf8]{inputenc}
\usepackage{subcaption}
\usepackage{amsmath,amssymb,amsfonts, bm}

\usepackage[linesnumbered,lined,boxed,commentsnumbered, ruled]{algorithm2e}
\usepackage{graphicx}
\usepackage{makecell}
\usepackage{multirow}
\usepackage{cite}
\usepackage{fouriernc}
\DeclareMathAlphabet{\mathcal}{OMS}{zplm}{m}{n}
\usepackage[table,xcdraw]{xcolor}
\usepackage{textcomp}
\usepackage{lipsum}
\usepackage{comment}
\usepackage{csquotes}
\usepackage{array}
\usepackage{xcolor}
\usepackage{url}
\usepackage{diagbox}
\usepackage[nameinlink,capitalise]{cleveref}
\usepackage{enumitem}
\def\BibTeX{{\rm B\kern-.05em{\sc i\kern-.025em b}\kern-.08em
    T\kern-.1667em\lower.7ex\hbox{E}\kern-.125emX}}

\usepackage[acronym,toc,nonumberlist]{glossaries}
\makeglossaries
\newacronym{5G}{5G}{Fifth Generation}
\newacronym{MIMO}{MIMO}{Multiple Input - Multiple Output}
\newacronym{SISO}{SISO}{single-input-single-output}
\newacronym{C-RAN}{C-RAN}{Centralized Radio Access Network}
\newacronym{TDM}{TDM}{time division multiplexing}
\newacronym{CoMP}{CoMP}{Coordinated Multi-Point}
\newacronym{RRU}{RRU}{remote radio unit}
\newacronym{BBU}{BBU}{baseband unit}
\newacronym{OPEX}{OPEX}{operating cost}
\newacronym{CAPEX}{CAPEX}{Capital cost}
\newacronym{CPRI}{CPRI}{Common Public Radio Interface}
\newacronym{eCPRI}{eCPRI}{eCommon Public Radio Interface}
\newacronym{OBSAI}{OBSAI}{Open base station architecture initiative}
\newacronym{Gbps}{Gbps}{Giga bits per second}
\newacronym{UE}{UE}{user equipmenFiber}
\newacronym{LTE}{LTE}{Long Term Evolution}
\newacronym{MMSE}{MMSE}{Minimum Mean Squared Error}
\newacronym{ZF}{ZF}{Zero Forcing}
\newacronym{LS}{LS}{Least Square}
\newacronym{BER}{BER}{bit error rate}
\newacronym{SNR}{SNR}{signal to noise ratio}
\newacronym{SINR}{SINR}{signal to interference and noise ratio}
\newacronym{PLC}{PLC}{Power-Line Communication}
\newacronym{AWGN}{AWGN}{additive white Gaussian noise}
\newacronym{PSD}{PSD}{power spectral density}
\newacronym{HARQ}{HARQ}{Hybrid automatic repeat request}
\newacronym{ARQ}{ARQ}{automatic repeat request}
\newacronym{NR}{NR}{New Radio}
\newacronym{PDU}{PDU}{power distribution unit}
\newacronym{FEC}{FEC}{forward error correction}
\newacronym{BW}{BW}{bandwidth}
\newacronym{CU}{CU}{Central Unit}
\newacronym{DU}{DU}{Distributed Unit}
\newacronym{RU}{RU}{Radio Unit}
\newacronym{AU}{AU}{Antenna Unit}
\newacronym{CSMA_CD}{CSMA/CD}{Carrier Sense Multiple Access/ Collision Detection}
\newacronym{SVD}{SVD}{singular value decomposition}
\newacronym{SR}{SR}{Selective repeat}
\newacronym{RS-FEC}{RS-FEC}{Reed-Solomon Forward Error Correction}
\newacronym{CRC}{CRC}{cyclic redundancy check}
\newacronym{C_M}{C\&M}{Control and Management}
\newacronym{NRZ}{NRZ}{non-return to zero}
\newacronym{PAPR}{PAPR}{peak to average power ratio}
\newacronym{FER}{FER}{frame error rate}
\newacronym{IND-Re}{IND-Re}{Impulsive noise detection/re-transmission}
\newacronym{uavs}{UAVs}{Unmanned Aerial Vehicles}
\newacronym{WSN}{WSN}{wireless sensor network}
\newacronym{MGDC}{MGDC}{Minimum Geometric Disk Cover}
\newacronym{TSP}{TSP}{Traveling Salesman Problem}
\newacronym{haps}{HAPS}{High Altitude Platform Station}
\newacronym{LoS}{LoS}{Line of Sight}
\newacronym{ILP}{ILP}{integer linear programming}
\newacronym{MILP}{MILP}{mixed integer linear programming}
\newacronym{INCM}{INCM}{IoT Node Clusters Minimization}
\newacronym{CM}{CM}{Cluster Minimization}
\newacronym{SCP}{SCP}{Set Covering Problem}
\newacronym{SCOP}{SCOP}{Set Covering Optimization Problem}
\newacronym{leach}{LEACH}{Low-Energy Adaptive Clustering Hierarchy}
\newacronym{IoT}{IoT}{Internet of Things}
\newacronym{SC}{SC}{Segment Clustering}
\newacronym{CETSP}{CETSP}{Close Enough Traveling Salesman Problem}
\begin{document}

\title{A Novel Framework of $K$-repetition Grant-free Access via Diversity Slotted Aloha (DSA)}

\author{\IEEEauthorblockN{ 
Haoran Mei\IEEEauthorrefmark{1},
Limei Peng\IEEEauthorrefmark{1},
and Pin-Han Ho\IEEEauthorrefmark{4}~\IEEEmembership{IEEE Fellow}, 
}\\
\IEEEauthorblockA{
    \IEEEauthorrefmark{1} 
    School of Computer Science and Engineering, 
    Kyungpook National University, 
    Deagu, South Korea\\
    }
\IEEEauthorblockA{
   \IEEEauthorrefmark{4} 
   Department of Electrical and Computer Engineering,
   University of Waterloo,
   Waterloo, ON, Canada \\}
\IEEEauthorblockA{Email: 
    \IEEEauthorrefmark{1}\{meihaoran, auroraplm\}@knu.ac.kr        
    \IEEEauthorrefmark{4}p4ho@uwaterloo.ca
   }
}

\maketitle
\begin{abstract}
This article introduces a novel framework of multi-user detection (MUD) for K-repetition grant-free non-orthogonal multiple access (K-GF-NOMA), called $\alpha$ iterative interference cancellation diversity slotted aloha ($\alpha$-IIC-DSA). The proposed framework targets at a simple yet effective decoding process where the AP can intelligently exploit the correlation among signals received at different resource blocks (RBs) so as to generate required multi-access interference (MAI) for realizing the signal-interference cancellation (SIC) based MUD. By keeping all operation and hardware complexity at the access point (AP), the proposed framework is applicable to the scenarios with random and uncoordinated access by numerous miniature mMTC devices (MTCDs). Numerical experiments are conducted to gain deep understanding on the performance of launching the proposed framework for K-GF-NOMA.
\end{abstract}

\begin{IEEEkeywords}
K-repetition grant-free non-orthogonal multiple access (K-GF-NOMA), massive machine type communication (mMTC), iterative interference cancellation (IIC), diversity slotted aloha (DSA).
\end{IEEEkeywords}

\section{Introduction}
Grant-free access (GFA) has been defined under 3GPP aiming to support uplink (UL) access over the 5G and beyond (B5G) massive machine-type communication (mMTC) scenario, which is featured by massive connectivity, low data rates, limited costs of devices, small-size packets, and sporadic and UL-biased traffic. As a random multiple access framework, GFA does not require the mMTC devices (MTCDs) to go through the conventional resource grant process, and thus can save most of the signaling overhead and access latency. Such advantage, nonetheless, is at the expense of possible collisions among the access attempts of multiple MTCDs for common resource blocks (RBs) due to the incongruous and uncoordinated resource selection at each distributed MTCD. 

To ensure reliable transmission in GFA, 3GPP has defined a two-step random access channel (RACH) procedure. This procedure uses message A (MsgA) to carry the preamble and payload signals in the UL, and message B (MsgB) for the random access response (RAR) and contention resolution in the downlink (DL) broadcast\cite{k-repetition}\cite{2-rach}. Specifically, an MTCD determines whether an UL access attempt made in the previous time frame was successful by reading the received MsgB. A retransmission is required if its ID is not found in MsgB.

Compared to the four-step RACH procedure typically used in grant-based access (GBA), the two-step RACH in GFA experiences significantly less signal overhead and access latency due to the simplified random access process, making it much more applicable to mMTC where massive miniature MTCDs are in place. However, this advantage comes at the expense of potential collisions among the uncoordinated access attempts of different MTCDs for common resource blocks (RBs), which may cause system outages.


\begin{figure*}[ht]
    \centering
    \includegraphics[width=0.95\textwidth]{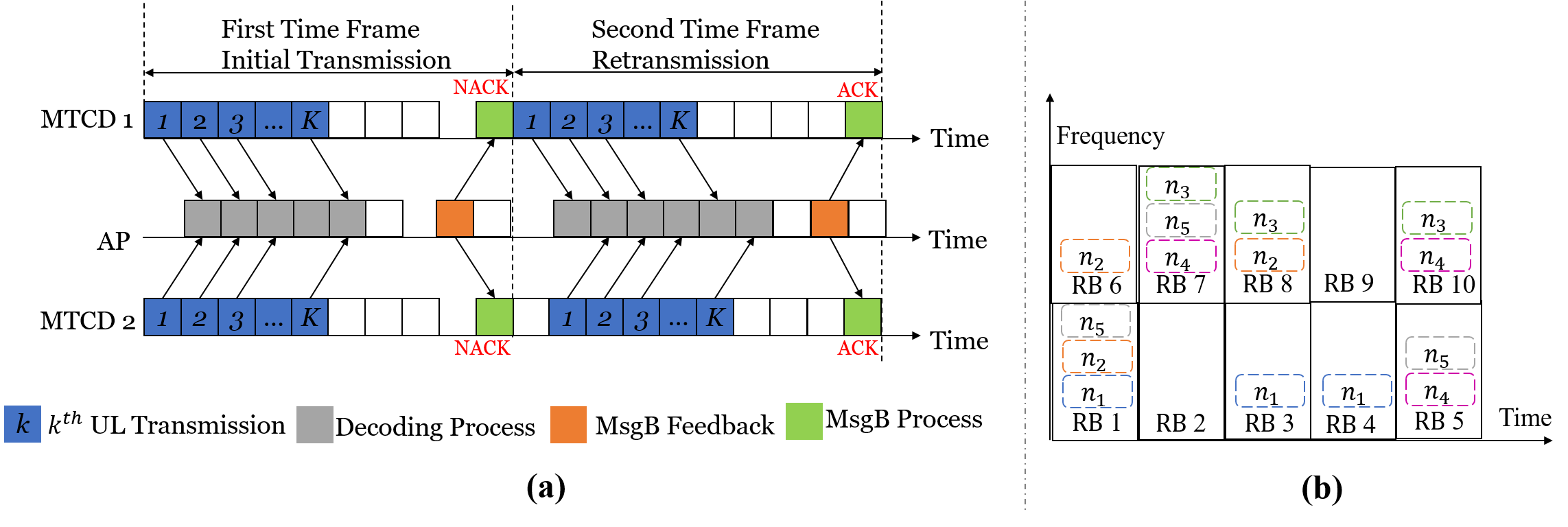}
    \caption{Standard signaling of K-GFA (a) K-Repetition Grant-Free Transmission Procedure (b) resource structure of a time frame and MTCDs distribution over RBs ($R$ = 10, $N$ = 5, $K$ = 3).}
    \label{fig: Fig1}
\end{figure*}

As a remedy, 3GPP defined $K$-repetition GFA (K-GFA) \cite{k-repetition} by launching multiple replicas of common packet within a time frame. Fig. \ref{fig: Fig1}(a) shows the transmission procedures of two MTCDs under K-GFA, where each square represents a time slot in a time frame. Specifically, each MTCD transmits a packet $K$ times to the access point (AP), with the receiver independently decoding each replica. Without loss of generality, each replica of a packet could be transmitted via any RB within a common time frame, and successful delivery of a packet requires at least one of the $K$ replicas to be successfully received and decoded by the AP. After all the received replicas are processed, the MTCDs then access the MsgB from the AP via a broadcast channel to determine whether the feedback is ACK or NACK. If an MTCD fails to transmit a packet to the receiver, it may retransmit the packet in the subsequent time frame.

K-GFA, incorporated with non-orthogonal multiple access (NOMA) and also referred to as K-GF-NOMA, is considered an interesting combination for which the corresponding techniques and design premises form a graceful complement to each other. To take the best advantage of K-GF-NOMA, implementing an effective multi-user detection (MUD) scheme at each RB for resolving the uncoordinated access attempts by the massive MTCDs is essential for achieving high-fidelity multiple access. Various MUD strategies are considered for K-GF-NOMA, such as power-domain, code-based, compressed sensing, and machine learning-based approaches. 


Different from the conventional MUD approaches, diversity slotted ALOHA (DSA)\cite{DALOHA} and its variants \cite{CRDSA1}\cite{CRDSA2} are allowed to explore signal-level correlation among RBs, where the user signal obtained from one RB can be used to facilitate the MUD of the others. With such an additional dimension of collision resolution, DSA can achieve significantly improved performance against its counterparts and is considered a promising candidate for the K-GF-NOMA. 

Motivated by the timeliness of this emerging topic, this article incorporates K-GF-NOMA with DSA by investigating a novel DSA-based MUD framework, called $\alpha$-iterative interference cancellation DSA ($\alpha$-IIC-DSA). The proposed $\alpha$-IIC-DSA iteratively performs successive interference cancellation (SIC) based MUD on the RBs in parallel, where the set of multi-access interference (MAI) used in an iteration is created by using the already obtained MTCD signals in the previous iterations. We introduce a general implementation model of the proposed $\alpha$-IIC-DSA MUD framework with its key parameters defined accordingly. Extensive simulation results help gain deep understanding on the access probability performance and the inferences/impacts of each key parameter.

The article is organized as follows. Section II provides a literature review. Section III presents a general implementation model of the proposed $\alpha$-IIC-RSA framework. Section IV presents the simulation results. Section V concludes the article.

\section{Literature Review}
\subsection{Collision resolution for GFA}
The state-of-the-art research on MUD for GF-NOMA falls in the following four categories, namely power domain GF (PD-GF) \cite{pdnoma1}\cite{pd-mud3}, code-based grant-free (CB-GF) \cite{scma_gf1}\cite{cd-based4}, compressed sensing (CS) \cite{cs-based1} \cite{cs-based3}, and machine learning (ML) \cite{ml-based1}\cite{ml-based2}.

PD-GF-NOMA focuses on performing MUD via SIC on each individual RB, where power level differences of superimposed signals are exploited. \cite{pdnoma1} concludes that precise power allocation is necessary for the AP to achieve desired system performance. \cite{pd-mud3} incorporates NOMA with multichannel ALOHA to enhance throughput of random access systems, where each MTCD transmits a packet with a power level based on its channel status indicator (CSI) and targeted received power level from a predefined pool. The authors developed an analytical model to establish the lower bound of the system's throughput, taking into account the given number of sub-channels and power levels.


Although feasible, PD-GF-NOMA becomes ineffective in the presence of a large number of MTCDs, where achieving a feasible distribution of power levels for signals superimposed at each RB becomes increasingly unlikely. Some research has resorted to CB-GF. For example, sparse code multiple access (SCMA) can integrate very well with GF-NOMA, thanks to its superb capability in handling a large and uncertain number of accessing users via overloaded codebooks, without the need for resource allocation or granting. In an SCMA system, each MTCD utilizes assigned codebooks instead of a traditional modulation scheme like quadrature phase shift keying (QPSK) for modulation. Here, one of the codewords from its codebook is selected for the transmission of an SCMA block consisting of a fixed number of RBs. Taking advantage of codeword sparsity, multiple MTCDs on a common SCMA block can be separated at the AP by using message passing algorithm (MPA), which computes the probabilities of all possible codeword combinations from the received signals and identifies the one with the maximum likelihood.

The vast number of MTCDs poses challenges related to codebook collisions. \cite{scma_gf1} investigates the impact of codebook design on the average symbol error probability (ASEP) of a GF-SCMA system experiencing codebook collisions. The study found that employing denser codebooks can enhance the robustness of ASEP, particularly under conditions of light codebook collisions and high signal-to-noise ratios.

On the other hand, while using more codes allows the system to accommodate a greater number of devices, it also results in increased MUD complexity. Consequently, there is a trade-off between the number of users the system can support and the complexity of codebook deployment at the receiver \cite{cd-based4}, and the resultant MUD performance could be significantly affected by environmental noise and interference.
This implies that CB-GF may not be suitable for scenarios involving a huge number of miniature MTCDs with limited capacity. 

CS-based MUD has been considered in GFA, thanks to the sparse user activity characteristic of mMTC. It is feasible to integrate CS with MPA to perform active user and data detection as a whole \cite{cs-based1}. \cite{cs-based3} introduced a compressive sampling matching pursuit (CoSaMP) algorithm for detecting non-zero elements in a signal, which can be applied to carry out active MTCDs detection based on the sporadic transmission principle. However, in general, CS-based MUD inherits the limitations of CB-GF and thus may only be applicable to certain scenarios.

Note that all the above MUD schemes are subject to dramatically increased computational complexity as the number of devices increases. Accordingly, some researchers have turned to ML-based approaches to achieve more efficient and robust results on MUD. \cite{ml-based1} employed cross-validation to determine the user's sparsity level, so the system does not have to predetermine it. \cite{ml-based2} applied deep learning to obtain a mapping between the received signal and active users, and the simulation results showed its significantly higher computation capability compared with conventional MUD algorithms. Although ML-based approaches can achieve efficient parameter estimation, such as accurate channel estimation and sparsity level estimation, model training may not always be feasible due to the lack of labeled data.

\subsection{Diversity Slotted ALOHA (DSA)}
Another dimension of contention resolution is via $K$-repetition by launching multiple copies of a packet on different RBs, where the success of an access attempt holds if any copy of the packet is received correctly. In this case, higher reliability is achieved via packet-level diversity. DSA is an initiative to enhance the performance of random access channel by using burst repetition\cite{DALOHA}. Copies of a common packet are launched either simultaneously on different frequency channels or spaced apart by random time slots on a single channel. Repetition patterns are designed upon frequency-division multiple access (FDMA) and time-division multiple access (TDMA), respectively. It was concluded that access probability can be improved by increasing the number of transmission attempts under sparse bursty traffic \cite{DALOHA}\cite{CRDSA1}.

In environments where the near-far effect is absent and NOMA offers no advantage, \cite{CRDSA1} introduces an advanced DSA upon TDMA system, known as contention resolution diversity slotted ALOHA (CRDSA), by employing SIC cross time slots. 
In each copy of a packet, a replica pointer is inserted to indicate the time slots used by the other copies of the same packet. It assumes that a user signal is decodable only if the signal is exclusive at a ''clean" time slot without collision with another. The channel information required for interference cancellation (IC) is obtained from the ''clean" time slot using a preamble (e.g., a binary sequence). Once a copy of a packet is decoded, the AP can know the time slots accessed by the other copies of the packet by reading the replica pointer carried in the copy and then perform SIC on those time slots accordingly. 

\cite{CRDSA2} manipulates the CRDSA framework and introduces irregular repetition slotted ALOHA (IRSA). IRSA incorporates randomness into the repetition times for each user, allowing the optimization of system performance by adjusting the probability distribution of repetition rates. This results in a solid reduction in packet loss probability compared to the other DSA counterparts. The paper provides a detailed description on the operational procedures using bipartite graphs, based on which an in-depth analysis and optimization of the normalized capacity of IRSA/CRDSA-based random access systems are conducted.

Over the years, random access protocols have evolved from DSA to more sophisticated protocols involving repetition diversity and NOMA, aiming to resolve more packet collisions by creating an additional power domain for multiplexing. For example, \cite{CRDSA4, pdnm_crdsa2} incorporates NOMA with CRDSA and IRSA, resulting in the CRDSA-NOMA system and IRSA-NOMA system, respectively. In these systems, packet replicas are transmitted with predetermined power levels using channel 
inversion at the MTCDs. This allows SIC within a single RB initially, followed by spanning other relevant RBs within a predefined time window. Built on the analytical model by \cite{CRDSA2}, these works analyze and optimize the packet loss probability of the CRDSA-NOMA and IRSA-NOMA system, respectively, under changing traffic loads to gain better understanding of the resultant system performance.

\section{Proposed $\alpha$-IIC-DSA Framwwork}
The previously reported DSA-based methods require some modifications of the standard at the MTCDs for localization of the packet copies. This may not be acceptable in some scenarios where the miniature MTCDs cannot provide the replica pointers. As a complement to the existing schemes, the proposed $\alpha$-IIC-DSA is characterized as a generic implementation of the DSA decoding mechanism while keeping the MTCDs completely transparent to all the add-on features at the AP. 

We consider a set of $N$ single-antenna MTCDs, each sending a single packet to the AP within each UL time frame following an arrival-and-go manner without coordination by the AP. Each UL time frame includes a number of $R$ RBs and $L$ predefined received power levels, and each MTCD sends a number of $K$ copies of its packet using randomly selected $K$ different RBs with a selected power level. Given an access threshold $\tau$ defining a tagged signal-to-interference-plus-noise ratio (SINR) for a successful access, the available power levels is determined as follows:
\begin{equation}
P_i = \tau (\sum^{i-1}_{j = 1}P_j + N_{0})
\end{equation}
where $P_i$ represents the $i$-th power level and $N_{0}$ refers to the additive white Gaussian noise (AWGN) with zero mean and variance $\sigma^2$. An example of a user access map is provided in Fig. \ref{fig: Fig1}(b), where 5 users denoted as $n_1, n_2, n_3, n_4$, and $n_5$ access the AP with 10 RBs and $K=3$. Due to random resource selection at each MTCD, a RB could be accessed by multiple MTCDs, potentially leading to collisions.

\subsection{System Description}
\begin{figure}[t]
    \centering
     \centering
     \includegraphics[width=0.49\textwidth]{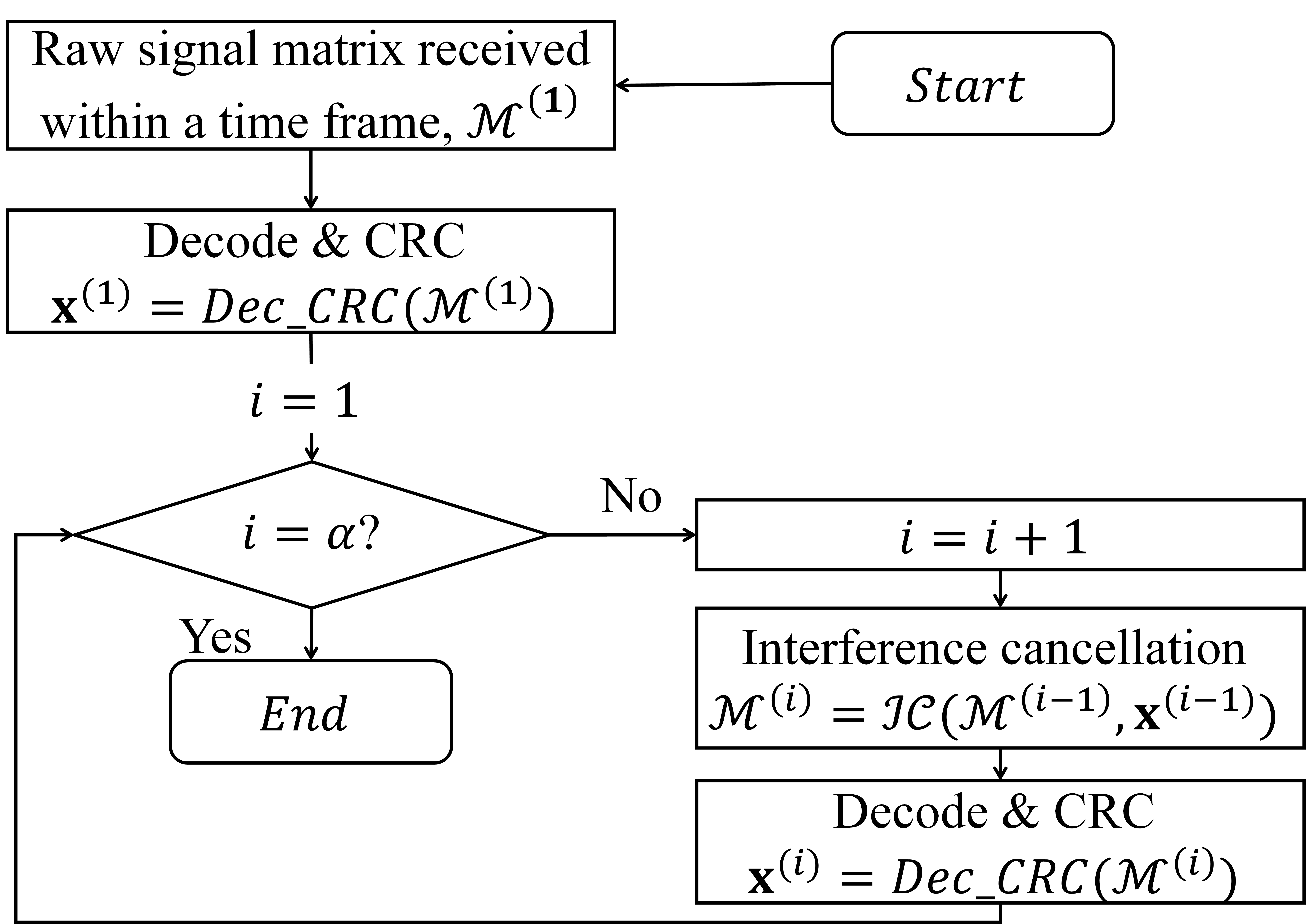}         
    \caption{Flow chart of the decoding process for the proposed $\alpha$-IIC-DSA framework.}
    \label{fig: diagram of IIC}
\end{figure}
\begin{figure*}[t]
    \centering
     \centering
     \includegraphics[width=0.71\textwidth]{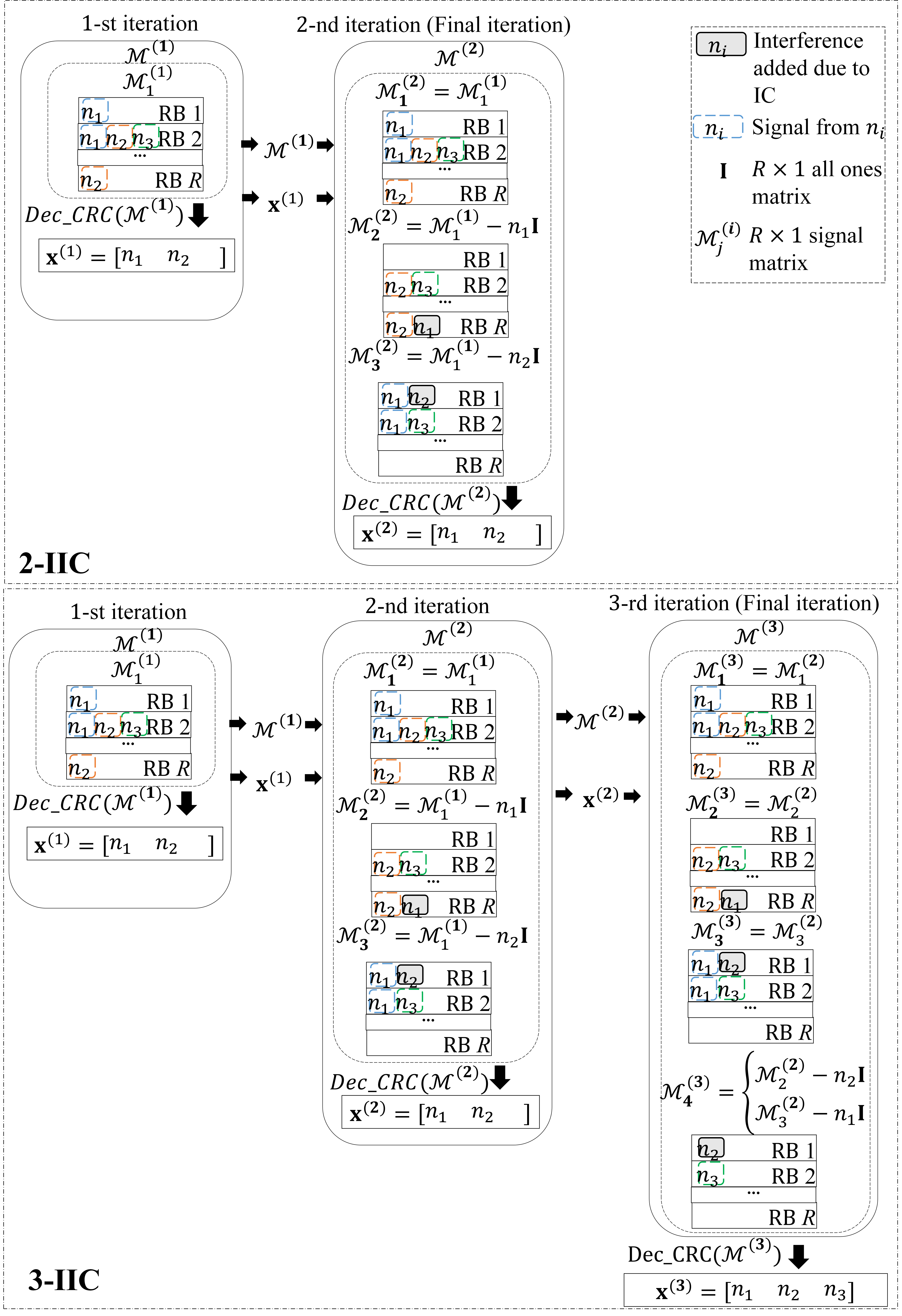}         
    \caption{The examples of the proposed $\alpha$-IIC-DSA module when $\alpha$ = 2, 3.}
    \label{fig2}
\end{figure*}

Similar to existing works \cite{CRDSA4}\cite{pdnm_crdsa2}, we assume perfect CSI is available at the AP to the proposed IC mechanism. we assume an MTCD signal of an RB is decodable if and only if the SINR of signal larger than a threshold $\tau$. Thus in the event that two or more signals contained in a single RB, whether a signal of the RB is decodable or not depends on the power level distribution of the signals. We also assume that the signal contained in an \emph{exclusive RB} without collision with another, such as $RB$ $3$, $RB$ $4$ and $RB$ $6$ in Fig. \ref{fig: Fig1}(b), must be with an SINR larger than $\tau$ and thus is sufficiently decodable. In this case the MTCD corresponding to any \emph{exclusive RB} is called an \emph{exclusive MTCDs}. 

The proposed $\alpha$-IIC-DSA framework is deployed at the AP to decode the multi-access signals of each RB in a time window. Assuming replica pointers are unavailable, the AP is unaware of which MTCD signals are contained in each RB, and the proposed framework implements \emph{iterative blind IC} in which all successfully obtained MTCD signals in the current iteration are buffered in a set of refined signal matrices and taken as MAIs for the IC process of the next iteration. 

\subsection{A General Model}
The proposed $\alpha$-IIC-DSA framework has two main function blocks executed in each iteration, namely interference cancellation ($\mathcal{IC}$) and decoding-CRC (\emph{Dec\_CRC}), respectively, as shown in Fig. \ref{fig: diagram of IIC}. The input of the $i$-th iteration consists of $\mathcal{M}^{(i-1)}$ and $\mathbf{x}^{(i-1)}$, which are the signal matrices as a result of IC and the matrix containing all successfully obtained MTCD signals up to the $(i-1)$-th iteration, respectively. The output of the iteration is denoted as $\mathcal{M}^{(i)}$ and $\mathbf{x}^{(i)}$, which is obtained via a set of operations defined as follows.

\textit{Definition 1}: The IC function, denoted as $\mathcal{IC}(\mathcal{M}^{(i-1)},\mathbf{x}^{(i-1)}$), bears the properties as follows:
\begin{itemize}
\item The output is a set of signal matrices, denoted as $\mathcal{M}^{(i)}$, where the elements are inherited from $\mathcal{M}^{(i-1)}$ and produced by the interference cancellation on the $\mathcal{M}^{(i-1)}$ with $\mathbf{x}^{(i-1)}$.;
\item $\mathbf{x}^{(0)} = \left\{0\right\}$ and $\mathcal{M}^{(0)}$ contains a single raw signal matrix from RBs;
\end{itemize}

\textit{Definition 2}: The decoding function $Dec\_CRC$($\mathcal{M}^{(i)}$) has the following properties:
\begin{itemize}
\item The output is a set of signals each being successfully decoded and validated via cyclic redundancy check (CRC), denoted as $\bold{x}^{(i)}$, where $|\bold{x}^{(i)}| \leq N + 1$;
\item $\bold{x}^{(i)}_{q}$ denotes the $q$-$th$ decoded signal of the $i$-$th$ iteration.
\end{itemize}

Specifically in the $i$-$th$ iteration, IC is performed on each signal matrix in $\mathcal{M}^{(i-1)}$, denoted as $\mathcal{M}^{(i-1)}_{j}$, by subtracting the MAI signal $\bold{x}^{(i-1)}_q$ from all elements within $\mathcal{M}^{(i-1)}_{j}$. Decoding and CRC are performed on each of the residual signal matrices, and the outcomes are stored in $\bold{x}^{(i)}$. Finally, $\bold{x}^{(i)}$ is taken as the MAI and fed into the IC function in the next iteration. The iterative process terminates if any of the following conditions is met: (1) the maximum number of iterations (i.e., $\alpha$) is reached, (2) all the MTCDs are recovered, and (3) no new MTCD is recovered and all possible aggregations of decoded signals from the previous iteration are used.

The parameter $\alpha$ dominates the decoding capacity of the proposed scheme. Consider $n_3$ in Fig. \ref{fig2}; initially, the signal cannot be obtained with $\alpha = 2$ because RB $2$ in $\mathcal{M}^{(2)}_{2}$ and $\mathcal{M}^{(2)}_{3}$, that contain the signal $n_3$, are still subject to collisions. On the other hand, taking $\alpha=3$ enables the successful decoding of $n_3$. In this case, it firstly decodes $n_1$ and $n_2$ from RB $1$ and RB $R$ in the first iteration. Once the signals of $n_1$ and $n_2$ are removed from RB $2$ in the second and third iterations, it comes up with a clean RB that exclusively contains the signal of $n_3$ in $\mathcal{M}^{(3)}_{4}$ that thus can be obtained by the end of the third iteration. 



To sum up, the proposed $\alpha$-IIC-DSA framework leverages buffered MTCD signals and make them iteratively decoded. Specifically, the memory is used to store the outcome sets, namely $\mathcal{M}^{(i)}$ and $\bold{x}^{(i)}$, one iteration after the other, where the buffered outcomes of an iteration are loaded for the subsequent iteration.

\subsection{Complexity Analysis}
With the proposed framework using blind IC, the storage and computation complexities exponentially increase as the system scales up. 
We formulate the asymptotic complexity of the blink IC process employed in the proposed $\alpha$-IIC-DSA framework by considering the following three factors: the number of memory write-read, the number of IC-decoding operations, and the storage usage. In addition to the case of $\frac{N}{2} \gg \alpha$, we also look into the special cases when $\alpha$ = 2 and $N$ as follows.
 
\subsubsection{$\alpha=2$}
In this case, the worst case complexity takes 2 iterations to terminate, leading to the complexity as follows:
\begin{itemize}
\item $1$-$st$ iteration: $N-1$ \textit{exclusive MTCDs} are recovered from $\mathcal{M}^{(1)}_{1}$, contributing to $N-1$ MAI signals.
\item $2$-$nd$ iteration: $N-1$ signal matrices are generated by IC with $N-1$ MAI signals on the matrix $\mathcal{M}^{(1)}_{1}$, where the remaining MTCD can be recovered.
\end{itemize}
During the whole process, only a received signal matrix and $N-1$ MAI signals are buffered. Thus, the storage complexity is expressed as $O(R+N)$. The computational complexity for memory write-read and decoding can be formulated as $O(2NR)$ and $O(NR)$, respectively.

\subsubsection{$\alpha$ for $\frac{N}{2} \gg \alpha$} We assume that $\frac{N}{2} \gg \alpha$. 
The worst case takes $\alpha$ iterations to terminate, leading to the complexity as follows:
\begin{itemize}
\item $1$-$st$ iteration: $N-2$ \textit{exclusive MTCDs} are decoded from $\mathcal{M}^{(1)}_1$, contributing to $N-2$ MAI signals.
\item $i$-$th$ iteration $(2 \leq i \leq \alpha)$: ${N - 2 \choose i - 1}$ different signal matrices are generated by the IC, and no new MTCD is decoded.
\end{itemize}
In summary, the amount of buffered signal matrices and MAI signals is no more than: $\sum^{\alpha-1}_{b = 1} {N - 2 \choose b - 1}$ and $N$, respectively, leading to the storage complexity expressed as $O(N^{(\alpha - 1)}R)$, while the computational complexity of memory write-read and decoding is $O(2\alpha N^{\alpha}R)$ and $O(\alpha N^{\alpha}R)$, respectively.

\subsubsection{$\alpha=N$} 
In this case, the worst case complexity in terms of required storage usage involves $N-1$ iterations, where each iteration is detailed as follows
\begin{itemize}
\item $1$-$st$ iteration: 1 \textit{exclusive MTCD} is recovered.
\item $i$-$th$ iteration ($2 \leq i \leq N-2$): $2^{(i-1)}$ signal matrices are generated from the IC and the number of decoded signals and MAI signals both increase by 1.
\item $(N-1)$-$th$ iteration: $2^{(N-2)}$ signal matrices are generated from the IC, and no new \textit{exclusive MTCDs} is decoded.
\end{itemize}
Moreover, the worst case complexity involves $N-2$ iterations detailed as:
\begin{itemize}
\item 1-st iteration: $N-2$ \textit{exclusive MTCDs} are recovered from $\mathcal{M}^{(1)}_{1}$, contributing to $N-2$ MAI signals.
\item $i$-$th$ iteration ($2 \leq i \leq N-2$): ${N-2 \choose i-1}$ signal matrices are generated by the IC with $N-2$ MAI signals on the matrices in $\mathcal{M}^{(i-1)}$, and no new MTCDs are decoded.
\end{itemize}
In summary, the amount of buffered signal matrices and MAI signals is no more than $2^{N-2}$ and $N-1$, respectively, leading to the storage complexity expressed as $O(2^{N-1}R + N)$, while the computational complexities of memory write-read and decoding are $O(2^{N-1}R + N^2)$ and $O(2^{N-1}R)$, respectively.

\begin{figure}[t]
    \centering
    \includegraphics[width=0.45\textwidth]{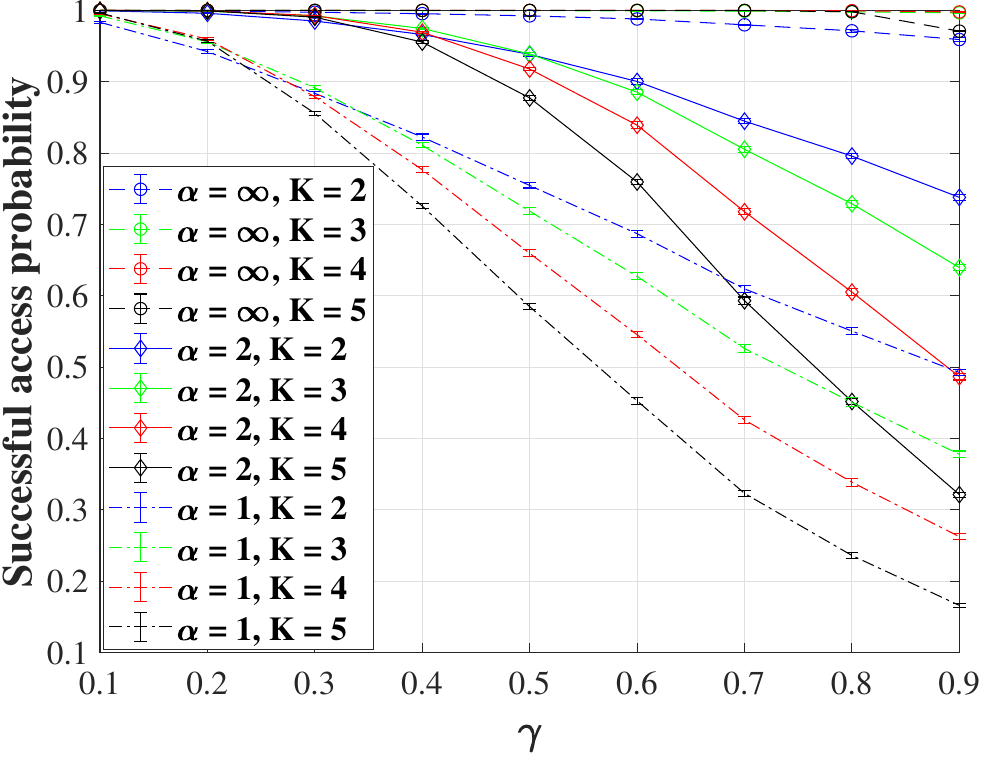}
    \caption{Expected access probability on $\alpha$-IIC-DSA system when $\alpha$ = 1, 2, $\infty$ and $K$ = 2, 3, 4, 5 for different $\gamma$.}
    \label{fig: Mresult1}
\end{figure}
\begin{figure}[tp]
    \centering
    \includegraphics[width=0.49\textwidth]{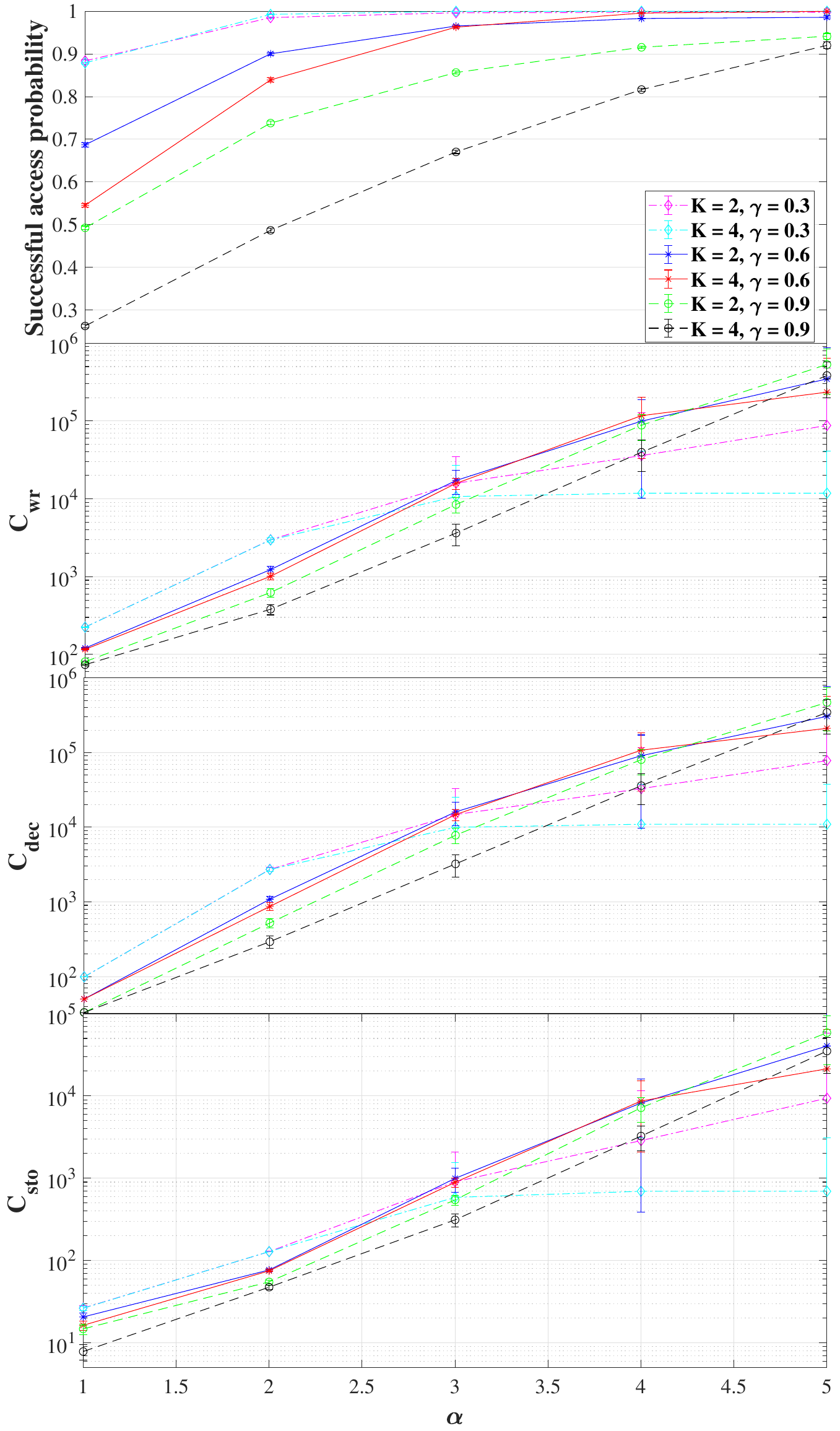}
    \caption{Expected access probability and complexities of the blind IC process when $K$ = 2, 4 and $\gamma$ = 0.3, 0.6, 0.9 for $\alpha$ ranged from 1 to 5, where $C_{wr}$, $C_{dec}$ and $C_{sto}$ denote average complexities for memory write-read, decoding and storage, respectively}
    \label{fig: Mresult2}
\end{figure}

\section{Case Study}
Numerical case studies are conducted to evaluate the access probability of the proposed $\alpha$-IIC-DSA framework with respect to a number of key parameters, including, $K$, $\alpha$ and user intensity $\gamma = \frac{N}{R}$. Assuming the access threshold $\tau$ is 1 and the noise power is fixed as 1 Watts, we define three available received power levels: 1, 2 and 4 Watts to form a power pool. In each time frame, every MTCD randomly selects $K$ RBs and a power level from the power pool for sending $K$ copies of the packet, and 10,000 time windows are examined.

Fig. \ref{fig: Mresult1} shows the access probability of the proposed $\alpha$-IIC-DSA framework versus the user intensity $\gamma$ ranged from 0.1 to 0.9 under $\alpha$ = 1, 2, $\infty$ and different values of $K$. It is observed that larger $\alpha$ and smaller $K$ generally leads to higher access probability. Specifically, there exists an optimal value of $K$ in each case; e.g., $K=2$ and $K=3$ for the case of $\alpha=1, 2$ and $\alpha=\infty$, respectively.

We also find that the user intensity $\gamma$ significantly affects the performance and the selection of operation parameters. For example, when $\alpha=2$, the ($\gamma$, $K$) tuple that can achieve the best performance is: ($\leq 0.3$, 3), ($\geq 0.4$, 2). From the operation perspective, the AP can estimate the current user intensity $\gamma$ according to its observed RB usage and inform each MTCD the best value of $K$ via MsgB in every time frame.

We further examine the cases with $\alpha$ ranging from 2 to 5 while taking $K$ = 2 and 4 for each case, and three user intensities: light ($\gamma$ = 0.3), moderate ($\gamma$ = 0.6), and heavy ($\gamma$ = 0.9), respectively. Fig. \ref{fig: Mresult2} shows the access probability as well as the average complexities, i.e., memory write-read operations ($C_{wr}$), decoding operations ($C_{dec}$), and storage space ($C_{sto}$), versus $\alpha$. 

Firstly, we find that the impact of $\alpha$ on access probability is more pronounced when $\gamma$ is large due to the fact that complete decoding is more likely to occur in the scenarios with smaller $\gamma$. Nonetheless, increasing $\alpha$ beyond a certain point yields rather marginal performance gain but meanwhile causing a substantial increase in complexity. As shown in Fig. \ref{fig: Mresult2}, over an order of magnitude on the increase of complexities in all the read-write, decoding, and storage operations is observed, while the improvement on the access probability is negligible when $\gamma \leq 0.6$. It is also the case when comparing the scenarios with $\alpha = 3$ and $\alpha = 4$ under $\gamma = 0.6$. Therefore, a proper trade-off between the access probability performance and system complexities should be identified by adjusting the value of $\alpha$ according to the value of $K$ and estimated user intensity $\gamma$, which is beyond the scope of this article and will be left to our future study.

\section{Conclusions}
The article introduced a novel K-GF-NOMA framework, namely $\alpha$-IIC-DSA, aiming to achieve effective contention resolution for uplink K-GFA in mMTC systems. The proposed framework is characterized by allowing dummy MTCDs without any knowledge on the add-on features at the AP, while the AP bears all the increased hardware and operation complexity. A general implementation model is provided as a detailed description of the proposed framework. Extensive case study disclosed the performance behaviour of the proposed framework with respect to a number of key parameters, including the number of iterations and traffic intensity, is investigated via extensive case study. We conclude that the proposed $\alpha$-IIC-DSA is a competitive candidate for the deployment of mMTC systems in 5G and Beyond communications.


\bibliographystyle{IEEEtran}
\bibliography{main.bib}

\end{document}